\documentclass[12pt,a4paper]{article}

\usepackage[a4paper,left=1in,right=1in,top=1in,bottom=1.15in,includeheadfoot]{geometry}
\usepackage[T1]{fontenc}
\usepackage[utf8]{inputenc}
\usepackage{lmodern}
\usepackage[doublespacing]{setspace}
\usepackage{microtype}
\usepackage{amsmath,amssymb,amsthm,mathtools,bm}
\usepackage{graphicx}
\usepackage{booktabs}
\usepackage[svgnames]{xcolor}
\usepackage{multirow}
\usepackage{tabularx}
\usepackage{array}
\usepackage[round,authoryear]{natbib}
\usepackage{url}
\usepackage[hidelinks]{hyperref}

\graphicspath{{images/}{Fig/}}

\DeclareMathOperator{\vech}{vech}
\DeclareMathOperator{\logit}{logit}
\DeclareMathOperator{\KL}{KL}

\DeclareMathOperator{\E}{\mathbb E}
\DeclareMathOperator{\TV}{TV}
\DeclareMathOperator{\diag}{diag}
\DeclareMathOperator{\offdiag}{offdiag}
\newcommand{\sigmoid}{\operatorname{logit}^{-1}}
\newcommand{\1}{\mathbf{1}}
\newcommand{\RR}{\mathbb R}

\newtheorem{theorem}{Theorem}
\newtheorem{lemma}{Lemma}

\newtheorem{corollary}{Corollary}

\theoremstyle{definition}

\newtheorem{assumption}{Assumption}

\theoremstyle{remark}

\title{A Bayesian Adaptive Latent Mixture Model for Zero-Inflated Weighted Brain Connectome Analysis}
\author{Hsin-Hsiung Huang\thanks{Corresponding author. Email: \href{mailto:Hsin-Hsiung.Huang@ucf.edu}{Hsin-Hsiung.Huang@ucf.edu}.}, Yuh-Haur Chen, and Teng Zhang\\
\small School of Data, Mathematical, and Statistical Sciences\\\small University of Central Florida, Orlando, Florida, USA}
\date{}

\begin{document}
\maketitle

\begin{abstract}
Replicated weighted networks often exhibit many structural zeros alongside heterogeneous non-zero edge strengths. In structural connectomics, this zero-inflation coincides with subjects expressing overlapping, rather than discrete, connectivity patterns. To address these features, we propose a Bayesian adaptive latent mixture model for zero-inflated weighted networks. Our approach represents each subject network as a simplex mixture of shared low-rank latent score matrices, integrated with a hurdle likelihood that separates edge existence from conditional edge strength. A sparsity-coupling parameter enables absent edges to be either independent of, or informative about, the latent connectivity. For computation, we employ transformed Hamiltonian Monte Carlo on unconstrained coordinates, selecting the number of templates via predictive fit, held-out link prediction, and template stability. Theoretically, we establish posterior consistency, local asymptotic normality, a Bernstein--von Mises approximation, and predictive consistency for an identifiable quotient-space estimand under a fixed-template scenario. Simulations demonstrate performance gains over topology-only baselines in settings with mixed memberships or structure-informed sparsity. Applied to Human Connectome Project data, the model recovers stable latent score patterns and heterogeneous subject-level mixtures, with behavioural analyses serving strictly as exploratory annotations rather than confirmatory biomarker claims.
\end{abstract}


\section{Introduction}
\label{sec:intro}

Multilayer and replicated networks serve as a primary mathematical framework for analysing relational data collected across multiple subjects, time points, or experimental conditions \citep{kivela2014multilayer,boccaletti2014structure}. In modern neuroimaging, a canonical instance is population connectomics, where each subject yields a connectivity matrix representing structural pathways or functional dependence among brain regions. Large consortia and standardised pipelines for diffusion MRI (dMRI) have made these replicated networks increasingly available, yet they present substantial statistical challenges because sparsity, measurement variability and cross-subject heterogeneity occur simultaneously \citep{vanessen2013hcp,glasser2013pipelines,Behrens2007,Cai2024}. A recurring scientific goal is to learn a small collection of shared population-level connectivity patterns and to quantify the extent to which each subject expresses them, thereby moving beyond deterministic clustering or single-network summaries.

Structural connectomes derived from dMRI are characterised by two pervasive features that make standard network models inadequate. First, the observed adjacency matrices are heavily zero inflated, where many edges are exactly zero because of biological absence, tractography uncertainty, or thresholding required for reproducibility \citep{Cai2024,Skoch2022}. Second, among the existing edges, the weights exhibit high variability and carry scientific meaning regarding the strength of neural communication. Standard approaches often treat zeros as small continuous weights or assume they are missing at random, which can lead to miscalibrated uncertainty for both topology and edge strength. This motivates a model that explicitly separates edge existence from conditional strength while allowing sparsity itself to be informative about the latent structure \citep{mullahy1986specification,simpson2015twopart}.

Existing methods such as ALMA \citep{fan2022alma} or the tensor block model \citep{wang2019multiway} typically impose discrete cluster assignments or orthogonal constraints. While computationally attractive, these constraints are often too restrictive for human brain organisation, where subjects express graded mixtures of shared connectivity motifs. To bridge these gaps, we propose the Bayesian Adaptive Latent Mixture Model (BALM). BALM represents each subject network as a simplex-constrained convex combination of shared low-rank latent templates on the logit-weight scale, explicitly modelling the zero-inflation through a hurdle architecture. In connectomics, that distinction is scientifically important because topological absence may arise partly from weak underlying pathways rather than from an entirely separate mechanism.

While motivated by connectomics, the BALM framework---specifically its hurdle-coupling mechanism and Dirichlet mixture architecture---is readily applicable to other networks characterised by zero inflation and weight heterogeneity, such as sparse social interaction patterns, financial transaction flows or large-scale logistical networks.

BALM makes several key contributions. Methodologically, it provides a fully generative Bayesian framework for sparse weighted networks with continuous mixed membership, yielding calibrated posterior uncertainty for topologies, weights and sparsity parameters via transformed Hamiltonian Monte Carlo. Theoretically, we establish posterior consistency, local asymptotic normality, a Bernstein--von Mises approximation and predictive consistency for an identifiable quotient-space estimand under a fixed-template scenario. Empirically, BALM recovers stable latent score patterns and heterogeneous subject-level mixtures in Human Connectome Project data, enabling exploratory behavioural annotations without requiring confirmatory biomarker claims.

The remainder of this article is organised as follows. Section~\ref{sec:model} introduces the model, Section~\ref{sec:posterior_inference} details computation, and Section~\ref{sec:theory} establishes asymptotic guarantees. Numerical simulations and the connectome application are presented in Sections~\ref{sec:numerical} and \ref{sec:hcp}, with all proofs and additional protocols deferred to the Supplementary Material.

\section{Model}
\label{sec:model}

The Bayesian Adaptive Latent Mixture (BALM) model offers a generative framework for replicated sparse weighted networks, designed for settings where subjects express multiple shared connectivity motifs simultaneously. This contrasts with optimisation-based approaches such as ALMA \citep{fan2022alma}, which target discrete layer-clustering by assuming each subject belongs to a single latent cluster. Rather than acting as a Bayesian wrapper around deterministic outputs, BALM enables principled statistical inference and uncertainty quantification for both network topology and continuous edge weights through a unified architecture.

To represent subject-level heterogeneity, BALM replaces discrete population partitions with simplex-constrained mixing weights $W_\ell \in \Delta^{M-1}$, allowing a continuous mixed-membership representation that better approximates graded brain organisation (Section~\ref{sec:model_templates}). Furthermore, it substitutes block-constant community structures with shared low-rank latent score matrices to capture diffuse connectome patterns. Finally, BALM incorporates a hurdle likelihood that explicitly separates edge existence from conditional edge strength. This formulation features an interpretable coupling parameter allowing the data to determine whether sparsity is informative about latent connectivity, a mechanism absent in topology-only baselines.

\subsection{Data and notation}
\label{sec:model_notation}

Let $A=\{A_\ell\}_{\ell=1}^L$ denote a collection of observed weighted adjacency matrices on a common set of $n$ nodes, where each layer $A_\ell\in[0,1]^{n\times n}$ is symmetric with zero diagonal. We assume weights have been scaled to lie in $[0,1]$, with a substantial fraction of exact zeros representing absent edges. After preprocessing, all nonzero weights satisfy $A_\ell(i,j)\in(0,1)$ so that the logit transform is finite. Let $\mathcal I=\{(i,j):1\le i<j\le n\}$ index the unique off-diagonal entries and define $P = |\mathcal I| = n(n-1)/2$.

For each layer $\ell$ and edge $(i,j)\in\mathcal I$, define
\[
Z_{\ell,ij}=\1\{A_\ell(i,j)>0\}, \qquad Y_{\ell,ij}=\logit\!\big(A_\ell(i,j)\big),
\]
with the convention that $Y_{\ell,ij}$ is only observed when $Z_{\ell,ij}=1$. We collect these into vectors $\mathbf Z_\ell\in\{0,1\}^P$ and $\mathbf Y_\ell\in\mathbb R^P$.

\subsection{Latent score templates and template-mixture representation}
\label{sec:model_templates}

BALM assumes that latent logit-scale edge strengths arise from a convex mixture of $M$ shared low-rank \emph{score} matrices. For $m=1,\dots,M$, let
\begin{equation}
\label{eq:model_S}
    S_m = \tau\,U_m \diag(\gamma_m)U_m^\top,
    \qquad
    U_m^\top U_m = I_K,
    \qquad
    \gamma_m\in\mathbb R^K,
\end{equation}
where $U_m\in\mathbb R^{n\times K}$ contains orthonormal latent basis vectors, $\gamma_m$ controls the signed magnitude of each latent mode in logit space, and the global scale $\tau>0$ separates overall template magnitude from relative structure. Allowing the entries of $\gamma_m$ to take both positive and negative values provides flexibility to capture both assortative and disassortative latent connectivity patterns.

Because the data involve only off-diagonal entries, the likelihood depends on the off-diagonal template $ Q_m = \offdiag(S_m)=S_m-\diag\{S_m\}$, rather than directly on $S_m$. Let $\mathbf q_m=\vech(Q_m)\in\mathbb R^P$ be the half-vectorisation of $Q_m$. For each layer $\ell$, the latent logit-scale mean vector is
\begin{equation}
\label{eq:mu_def}
\boldsymbol\mu_\ell = \sum_{m=1}^M W_{\ell m}\,\mathbf q_m,
\qquad
W_{\ell m}\ge 0,\quad \sum_{m=1}^M W_{\ell m}=1.
\end{equation}
Equivalently, $\mu_{\ell,ij}=\sum_{m=1}^M W_{\ell m}\,q_{m,ij}$. This simplex mixture allows each subject network to express graded and overlapping combinations of shared connectivity patterns rather than being assigned to a single latent class.

\noindent\textbf{Identifiability and invariances.}
As in other mixture and factor models, the representation in \eqref{eq:model_S}--\eqref{eq:mu_def} is not globally identifiable without conventions. The likelihood is invariant to permutations of the template index. Within each template, the factorisation of the score matrix is invariant to orthogonal rotations within eigenspaces corresponding to repeated values of $\gamma_{mk}$, and to deterministic sign changes induced by equivalent spectral coordinates. Since the data use only off-diagonal entries, the diagonal of $S_m$ is not identifiable from the likelihood, which is why the estimand is taken to be $Q_m=\offdiag(S_m)$.

All theoretical statements are therefore made on a quotient space that identifies parameters up to template permutation and, locally, up to the internal spectral symmetries of the score matrices. For posterior summaries, we align templates across Markov chain Monte Carlo draws by solving a permutation matching problem that maximises correlation of the vectorised upper-triangular entries of $\{Q_m\}_{m=1}^M$, implemented via the Hungarian algorithm, and we report aligned posterior means and credible intervals for the off-diagonal templates and associated subject-specific weights.

\subsection{Hurdle likelihood with structure-informed sparsity}
\label{sec:likelihood}

In connectomics and related fields, observed edge weights are continuous values bounded in $[0,1]$ with a substantial fraction of exact zeros representing absent connections. Continuous distributions supported on $\RR$ cannot generate the $-\infty$ required by the logit transformation to represent an exact zero. A hurdle architecture is therefore natural in this setting: it explicitly separates topological absence, handled by a point mass at zero, from connection strength, handled by a continuous distribution on the strictly positive weights \citep{simpson2015twopart,pramanik2025hurdle}.

Given latent means $\boldsymbol\mu_\ell$, BALM defines the probability of edge presence for each layer $\ell$ and edge $(i,j)\in\mathcal I$ as
\begin{equation}
\label{eq:pi_def}
\pi_{\ell,ij}=\sigmoid\!\left(a_0+a_1\mu_{\ell,ij}\right),
\end{equation}
where $a_0$ controls global sparsity and $a_1$ governs coupling between latent strength and edge existence. The presence indicator follows
\begin{equation}
\label{eq:lik_Z}
Z_{\ell,ij}\sim\mathrm{Bernoulli}(\pi_{\ell,ij}).
\end{equation}
The parameter $a_1$ has a direct interpretation that is particularly useful for connectomes: $a_1=0$ yields a decoupled model in which sparsity is independent of latent connectivity, while $a_1\neq 0$ allows topological zeros to be informative about weak latent connectivity.

Conditional on $Z_{\ell,ij}=1$, the observed logit-scale weight satisfies
\begin{equation}
\label{eq:lik_Y}
Y_{\ell,ij}\mid Z_{\ell,ij}=1 \sim \mathcal N(\mu_{\ell,ij},\sigma^2).
\end{equation}
For robustness to outliers, a Student-$t$ likelihood may replace the Gaussian component, $Y_{\ell,ij}\mid Z_{\ell,ij}=1 \sim t_{\nu}(\mu_{\ell,ij},\sigma)$, with fixed degrees of freedom $\nu$. Assuming conditional independence across edges and layers, the likelihood factorises over $\ell=1,\dots,L$ and $(i,j)\in\mathcal I$.

\subsection{Prior distributions}
\label{sec:priors}

The prior respects the model constraints while remaining weakly informative and flexible enough to accommodate subject-level covariates. In the baseline model without covariates, layer-specific weights receive an exchangeable Dirichlet prior, $W_\ell \sim \mathrm{Dir}\{(\alpha/M)\mathbf 1_M\}$ for $\ell=1,\dots,L$, where $\alpha$ controls whether layers tend to be dominated by a few templates or are more uniformly mixed. When subject-level covariates $x_\ell \in \mathbb{R}^p$ are available, the mixing weights instead follow a covariate-driven logistic-normal distribution. We use a baseline-category parameterisation, setting $\psi_{\ell 1}=0$ and, for $m=2,\dots,M$, defining $\psi_{\ell m}=x_\ell^\top \beta_m + \epsilon_{\ell m}$ with $\epsilon_{\ell m} \sim \mathcal N(0,\sigma_\epsilon^2)$. The simplex weights are then
\[
W_{\ell m}=\frac{\exp(\psi_{\ell m})}{\sum_{k=1}^M \exp(\psi_{\ell k})},\qquad m=1,\dots,M.
\]
To prevent overfitting, we impose Gaussian shrinkage priors on the regression coefficients, $\beta_{mj} \sim \mathcal N(0,\sigma_\beta^2)$.

Each template score matrix is parameterised via \eqref{eq:model_S}. We use the QR-induced Stiefel construction, which induces a rotation-invariant prior on $U_m$ when the unconstrained matrix has independent standard Gaussian entries. Spectral weights receive independent Normal priors, $\gamma_{mk} \sim \mathcal{N}(0, \sigma^2_\gamma)$ for $k=1,\dots,K$ and $m=1,\dots,M$. A global scale parameter separates overall magnitude from relative structure, $\tau \sim \mathrm{HalfNormal}(\sigma_\tau)$. Sparsity parameters receive independent mean-zero Gaussian priors, $a_0 \sim \mathcal N(0,\sigma_{a_0}^2)$ and $a_1 \sim \mathcal N(0,\sigma_{a_1}^2)$, and the conditional logit-scale noise variance has a weakly informative inverse-Gamma prior, $\sigma^2 \sim \mathrm{Inv\text{-}Gamma}(a_\sigma,b_\sigma)$. Hyperparameters are treated as fixed throughout.

\section{Posterior computation and model selection}
\label{sec:posterior_inference}

Let $\theta = \big( W, \{U_m,\gamma_m\}_{m=1}^M, \tau, a_0, a_1, \sigma^2, \beta, \epsilon \big)$ denote the collection of all unknown parameters in BALM, where $\beta$ and $\epsilon$ are included only in the optional covariate-driven specification. The score matrices are constructed as $S_m = \tau U_m \diag(\gamma_m) U_m^\top$, the estimands of interest are $Q_m=\offdiag(S_m)$, and latent means satisfy $\mu_{\ell,ij}=\sum_{m=1}^M W_{\ell m}q_{m,ij}$. The observed data $A$ are deterministically transformed into $(Z,Y)$ via thresholding and logit mapping. Combining the hurdle likelihood with the prior distributions gives
\begin{align*}
p(\theta \mid A, X) &\propto p(Z \mid W,Q,a_0,a_1)\; p(Y \mid Z,W,Q,\sigma^2)\; \prod_{\ell=1}^L p(W_\ell \mid x_\ell, \beta)\\
&\quad \times \prod_{m=1}^M p(U_m)\,p(\gamma_m)\; p(\tau)\,p(a_0)\,p(a_1)\,p(\sigma^2)\,p(\beta).
\end{align*}
In the baseline model without covariates, $\beta$ and $\epsilon$ are omitted and $p(W_\ell \mid x_\ell, \beta)$ simplifies to the marginal Dirichlet prior $p(W_\ell)$.

\subsection{Constrained parameterisation for HMC}
\label{sec:constraints}

Due to the nonlinearity of the Bernoulli--logit component, the orthogonality constraints on $U_m$, and the simplex constraints on $W_\ell$, closed-form posterior expressions are unavailable. We therefore employ Hamiltonian Monte Carlo equipped with the No-U-Turn Sampler. To respect the parameter constraints within an unconstrained Euclidean sampling space, we apply specific transformations: additive log-ratio coordinates for the simplex weights $W_\ell$, log-transformations for positive scalars ($\tau$, $\sigma^2$), and a QR-induced Stiefel construction for the orthonormal matrices $U_m$.

The log-posterior evaluated by the sampler incorporates the necessary log-Jacobian adjustments for the bijective transformations. The full details of these unconstrained parameterisations, their inverse mappings, and the induced marginal distributions on the constrained space are provided in Appendix C.1 of the Supplementary Material.

\subsection{Adaptation, convergence and computation}
\label{sec:adaptation}

The sampler includes a warm-up phase to tune the step size and mass matrix. The leapfrog step size is adapted to achieve a target acceptance probability around $0.80$, while the mass matrix is estimated from warm-up draws to precondition the Hamiltonian dynamics. Convergence and sampling efficiency are assessed using rank-normalised potential scale reduction, bulk and tail effective sample sizes, and post-warm-up divergences \citep{vehtari2021rank}. Persistent divergences after warm-up are treated as evidence that reparameterisation or stronger priors may be required.

The cost of one gradient evaluation is dominated by constructing the dense score matrices $S_m$ and projecting them to layer-specific means. For $L$ layers, $n$ nodes, $M$ templates and template rank $K$, the per-evaluation complexity is
\[
\mathcal{O}\!\left(MKn^2 + LMP\right),\qquad P=n(n-1)/2.
\]
Thus BALM scales linearly with the number of layers and quadratically with the number of nodes, as expected for dense connectome matrices.

\subsection{Model selection and validation}
\label{sec:model_selection}

Selecting the number of templates $M$ is a critical step in identifying interpretable brain network states. Following established practice in Bayesian learning theory, we use a multi-criteria strategy combining predictive density assessment, held-out topology prediction and structural reproducibility \citep{watanabe2010asymptotic,ghasemian2019evaluating,brunet2004metagenes}.

\noindent\textbf{Widely applicable information criterion.}
Let $\{\theta^{(s)}\}_{s=1}^S$ denote post-warm-up posterior draws. For model comparison, we compute the edge-wise information criterion using the pointwise hurdle log likelihood contribution, defined as
\begin{equation}
\label{eq:pointwise_loglik}
\ell_{\ell,ij}(\theta) = \log p\!\left(Z_{\ell,ij}\mid \theta\right) + Z_{\ell,ij}\,\log p\!\left(Y_{\ell,ij}\mid Z_{\ell,ij}=1,\theta\right).
\end{equation}
Using this, we calculate the log pointwise predictive density and the effective number of parameters:
\[
\mathrm{lppd}_{\ell,ij}=\log\!\left(\frac{1}{S}\sum_{s=1}^S \exp\{\ell_{\ell,ij}(\theta^{(s)})\}\right), \qquad p_{\mathrm{waic},\ell,ij}=\mathrm{Var}_{s=1,\dots,S}\!\left(\ell_{\ell,ij}(\theta^{(s)})\right).
\]
Summing over all edges and layers yields the total lppd and penalty $p_{\mathrm{waic}}$, and $\mathrm{WAIC} = -2(\mathrm{lppd} - p_{\mathrm{waic}})$. This edge-wise criterion targets held-out edge prediction within observed layers. Prediction for a new layer instead corresponds to the collapsed layer likelihood in which the layer-specific weights are integrated out. In flexible latent models, WAIC can continue to decrease mildly with $M$, so it is not used in isolation.

\noindent\textbf{Held-out link prediction.}
To assess generalisation to unobserved structure, we randomly mask a fraction of the entries in the adjacency matrices, fit the model on the remaining entries, and compute posterior mean probabilities of edge existence for held-out edges:
\[
\hat{\pi}_{\ell,ij} = \frac{1}{S} \sum_{s=1}^S \sigmoid\left(a_0^{(s)} + a_1^{(s)} \mu_{\ell,ij}^{(s)}\right).
\]
We then compute the receiver-operating-characteristic area under the curve by comparing the predicted probabilities $\hat{\pi}_{\ell,ij}$ against the true held-out binary presence $Z_{\ell,ij}$.

\noindent\textbf{Consensus stability.}
A robust choice of $M$ should yield consistent templates across independent runs. For fixed $M$, we execute the sampler $R$ times with different random seeds. Let $\hat{Q}^{(r)}$ denote the posterior mean template tensor from run $r$. To account for label switching, we align runs $r$ and $r'$ by finding the optimal permutation implemented via the Hungarian algorithm that maximises pairwise correlation between templates. The stability score is defined as
\begin{equation}
\label{eq:stability}
\mathrm{Stability}(M) = \frac{1}{\binom{R}{2}} \sum_{r < r'} \frac{1}{M} \sum_{m=1}^M \rho\!\left(\hat{Q}_{m}^{(r)}, \hat{Q}_{\pi_{r,r'}(m)}^{(r')}\right),
\end{equation}
where $\rho(\cdot, \cdot)$ denotes the Pearson correlation of the vectorised upper-triangular entries. Higher stability indicates greater structural reproducibility.

\section{Theoretical study}
\label{sec:theory}

The asymptotic guarantees for BALM are established under a condition motivated by population connectomics. We fix the number of nodes $n$ (and hence the edge count $P=n(n-1)/2$ per layer) alongside the template complexity $(M,K)$, while the number of replicated networks $L$ grows. By treating the layer-specific weights $W_1,\dots,W_L$ as latent variables that are integrated out, the observed layer datasets become independent and identically distributed from a marginal distribution indexed by a finite-dimensional global parameter. All theoretical results are stated modulo label switching. The asymptotic guarantees established in this section apply specifically to the baseline model equipped with the exchangeable Dirichlet prior. While the covariate-driven logistic-normal extension shares identical layer-specific geometric properties, its theoretical analysis involves non-identically distributed layers and falls beyond the scope of the current fixed-distribution condition.

\subsection{Marginal layer model and quotient parameterisation}
\label{sec:theory_setup}

For each layer $\ell$, define the hurdle data $D_\ell := \big\{(Z_{\ell,ij},\, Y_{\ell,ij}\cdot \1\{Z_{\ell,ij}=1\}) : (i,j)\in\mathcal I\big\}$.
Let the global parameter be $\eta := \big(\{Q_m\}_{m=1}^M,\; a_0,\; a_1,\; \sigma^2 \big)$, where each $Q_m$ lies on the low-rank quotient manifold induced by $S_m=\tau U_m\diag(\gamma_m)U_m^\top$, rather than on the full ambient space of all off-diagonal matrices. Conditional on $(W_\ell,\eta)$ the likelihood factorises across edges. The collapsed likelihood for one layer is
\begin{equation}
    \label{eq:marg_layer_theory}
    p_\eta(D_\ell)=
    \int_{\Delta^{M-1}} p(D_\ell\mid w,\eta)\;
    \mathrm{Dir}\!\left(w;\tfrac{\alpha}{M}\mathbf 1_M\right)\,dw.
\end{equation}
Under the data-generating parameter $\eta_0$, the layers satisfy $D_1,\dots,D_L \stackrel{\mathrm{i.i.d.}}{\sim} p_{\eta_0}$.

Because the likelihood is invariant to permutations of template labels, we work on the quotient space. Write $\eta\sim\eta'$ if $(a_0,a_1,\sigma^2)=(a_0',a_1',\sigma'^2)$ and there exists a permutation $\pi\in\mathfrak S_M$ such that $Q'_m=Q_{\pi(m)}$ for all $m$. Define the permutation-invariant metric
\begin{equation}
\label{eq:d_eta_theory}
    d(\eta,\eta'):=
    \min_{\pi\in\mathfrak S_M}
    \left(
    \sum_{m=1}^M \|Q_m - Q'_{\pi(m)}\|_F^2
    \right)^{1/2}
    +
    |a_0-a_0'| + |a_1-a_1'| + |\sigma^2-\sigma'^2|.
\end{equation}

To apply local asymptotic arguments, we fix a local coordinate chart around $\eta_0$ on the quotient space. Since $\mathfrak S_M$ is finite and the true templates are separated, there exists a neighbourhood of $\eta_0$ in which the minimising permutation in \eqref{eq:d_eta_theory} is unique and equal to the identity. Within that neighbourhood, one may represent the equivalence class $[\eta]$ by a uniquely labelled representative and introduce a smooth local coordinate map $\phi(\eta)\in\mathbb R^{d_\eta}$.

\subsection{Assumptions}
\label{sec:theory_assumptions}

The score-matrix representation $S_m=\tau U_m \diag(\gamma_m)U_m^\top$ introduces internal rotational and sign invariances. By defining the quotient metric directly on the reconstructed off-diagonal templates $Q_m$ via the Frobenius norm, we quotient out these internal symmetries and treat $\{Q_m\}$ as the estimand. We make the following assumptions.

\begin{assumption}[Effective compact support via prior tails]
\label{ass:compact_theory}
For the asymptotic analysis, the effective parameter space is restricted to a compact region bounded by finite constants $B_Q,B_a$ and $0<\sigma_{\min}^2<\sigma_{\max}^2<\infty$ such that $\|Q_m\|_F\le B_Q$ for all $m$, $|a_0|\le B_a$, $|a_1|\le B_a$, and $\sigma_{\min}^2\le\sigma^2\le\sigma_{\max}^2$.
\end{assumption}

\begin{assumption}[Identifiability and template separation]
\label{ass:ident_theory}
If $p_\eta(\cdot)=p_{\eta'}(\cdot)$ almost everywhere, then $\eta\sim\eta'$. Moreover, the true templates are distinct with a strict separation margin $\Delta_0 := \min_{m\neq m'} \|Q_{0m}-Q_{0m'}\|_F > 0$. The underlying true score matrices possess simple non-zero spectra so that local spectral coordinates are defined after a deterministic sign convention.
\end{assumption}

\begin{assumption}[Smoothness, moment bounds, and nonsingular information]
\label{ass:smooth_theory}
Let $\phi=\phi(\eta)$ denote local coordinates on the quotient space in a neighbourhood of $\eta_0$, with $\phi_0=\phi(\eta_0)$. There exists an open neighbourhood $\mathcal N$ of $\phi_0$ such that for almost every $D_\ell$, the map $\phi\mapsto \log p_\eta(D_\ell)$ is three times continuously differentiable on $\mathcal N$ with third derivatives admitting an integrable envelope under $P_{\eta_0}$, and the one-layer Fisher information matrix $I(\phi_0):= -\mathbb{E}_{\eta_0}[\nabla_\phi^2 \log p_\eta(D_\ell)|_{\phi=\phi_0}]$ is finite and positive definite.
\end{assumption}

\begin{assumption}[Prior thickness in local coordinates]
\label{ass:prior_theory}
In the local coordinates $\phi$, the induced prior has a density $\pi(\phi)$ that is continuous at $\phi_0$ and satisfies $\pi(\phi_0)>0$.
\end{assumption}

For the Bernstein--von Mises result, we additionally use the local quadratic domination condition stated as Supplementary Assumption 5. This is a standard posterior-localisation condition in finite-dimensional regular parametric problems and is made explicit in the Supplementary Material to avoid overloading the main text.

\subsection{KL support and posterior consistency}
\label{sec:kl_support}

Let $\nu$ be a $\sigma$-finite dominating measure for $D_\ell$, for example a product of counting measure on $\{0,1\}^P$ for $Z$ and Lebesgue measure on $\mathbb R^P$ for a padded vector of $Y$ values. Write $p_\eta$ for the density of $D_\ell$ with respect to $\nu$. Define the Kullback--Leibler divergence $\KL(\eta_0,\eta) := \E_{\eta_0}[\log \{p_{\eta_0}(D_\ell)/p_\eta(D_\ell)\}]$.

\begin{lemma}[KL continuity and KL support]
\label{lem:kl_support_theory}
Under Assumptions~\ref{ass:compact_theory} and \ref{ass:prior_theory}, for every $\epsilon>0$ there exists $\delta>0$ such that $d(\eta,\eta_0)<\delta \Rightarrow \KL(\eta_0,\eta)<\epsilon$.
Moreover, the prior assigns positive mass to $\{\eta:\KL(\eta_0,\eta)<\epsilon\}$.
\end{lemma}

\begin{theorem}[Posterior consistency modulo permutation]
\label{thm:posterior_consistency_theory}
Under Assumptions~\ref{ass:compact_theory}--\ref{ass:prior_theory}, for every $\epsilon>0$,
\[
\Pi\big(d(\eta,\eta_0)>\epsilon \mid D_1,\dots,D_L\big)\to 0
\quad \text{in } P_{\eta_0}\text{-probability as } L\to\infty.
\]
\end{theorem}

\subsection{Local asymptotic normality and Bernstein--von Mises approximation}
\label{sec:lan_bvm}

Write the marginal log likelihood in local coordinates as $ \ell_L(\phi) := \sum_{\ell=1}^L \log p_\eta(D_\ell)$, $ \phi=\phi(\eta)$.
Let $ s_{\phi}(D_\ell) := \nabla_\phi \log p_\eta(D_\ell)$, $H_{\phi}(D_\ell) := \nabla_\phi^2 \log p_\eta(D_\ell)$, and define the normalised score at the truth by $\Delta_L := L^{-1/2}\sum_{\ell=1}^L s_{\phi_0}(D_\ell)$.

\begin{lemma}[LAN expansion]
\label{lem:lan}
Under Assumption~\ref{ass:smooth_theory}, for every fixed $B<\infty$,
\[
\sup_{\|h\|\le B}\left|
\ell_L(\phi_0+h/\sqrt{L})-\ell_L(\phi_0)
-
h^\top \Delta_L
+\frac12 h^\top I(\phi_0) h
\right|
\xrightarrow[L\to\infty]{P_{\eta_0}} 0.
\]
Moreover, $\Delta_L \Rightarrow \mathcal N(0,I(\phi_0))$ under $P_{\eta_0}$.
\end{lemma}

\begin{theorem}[Bernstein--von Mises in local coordinates]
\label{thm:bvm_theory}
Assume Assumptions~\ref{ass:compact_theory}--\ref{ass:prior_theory} and the local quadratic domination condition stated as Supplementary Assumption 5. Let $\Pi(\cdot\mid D_{1:L})$ be the posterior on $\phi$ induced by the prior on $\eta$ and the marginal likelihood \eqref{eq:marg_layer_theory}. Let $\mathsf N_L$ denote the Gaussian distribution on $\mathbb R^{d_\eta}$ with mean $I(\phi_0)^{-1}\Delta_L$ and covariance $I(\phi_0)^{-1}$. Then, under $P_{\eta_0}$,
\[
\left\|
\Pi\!\left(\sqrt{L}(\phi-\phi_0)\in \cdot \,\middle|\, D_{1:L}\right)
-
\mathsf N_L(\cdot)
\right\|_{\mathrm{TV}}
\to 0.
\]
\end{theorem}

\begin{corollary}[Parametric posterior contraction]
\label{cor:contraction}
Under the assumptions of Theorem~\ref{thm:bvm_theory}, for any sequence $M_L\to\infty$,
\[
\Pi\left(\|\phi-\phi_0\|>\frac{M_L}{\sqrt{L}} \,\middle|\, D_1,\dots,D_L\right)\to 0
\quad\text{in } P_{\eta_0}\text{-probability.}
\]
Equivalently, in the original quotient metric, for any $M_L\to\infty$,
\[
\Pi\left(d(\eta,\eta_0)>\frac{M_L}{\sqrt{L}} \,\middle|\, D_1,\dots,D_L\right)\to 0
\quad\text{in } P_{\eta_0}\text{-probability.}
\]
\end{corollary}

\subsection{Posterior predictive convergence and layer-specific weights}
\label{sec:pred_weights}

\begin{theorem}[Posterior predictive convergence]
\label{thm:pred_conv_theory}
Under Assumptions~\ref{ass:compact_theory}--\ref{ass:prior_theory}, let $p^{\mathrm{post}}(D_{\mathrm{new}}\mid D_{1:L}) = \int p_\eta(D_{\mathrm{new}})\, d\Pi(\eta\mid D_{1:L})$ be the posterior predictive distribution for a new layer. Then
\[
\TV\!\left(p^{\mathrm{post}}(\cdot\mid D_{1:L}),\; p_{\eta_0}(\cdot)\right)\to 0
\quad\text{in } P_{\eta_0}\text{-probability.}
\]
\end{theorem}

We next consider a separate asymptotic scenario relevant to estimating a single subject's weight vector when the number of nodes grows. Fix a layer $\ell$ and treat the edges $(i,j)\in\mathcal I$ as conditionally independent given $(W_\ell,\eta_0)$. Let $P=|\mathcal I|$ grow with $n$.

\begin{theorem}[Conditional contraction of $W_\ell$]
\label{thm:weights_theory}
Fix a layer $\ell$ and suppose the global parameter equals $\eta_0$. Assume there exists $w_0\in\mathrm{int}(\Delta^{M-1})$ such that the conditional model for $D_\ell$ is correctly specified at $W_\ell=w_0$. Assume the conditional log likelihood is twice continuously differentiable in $w$ on a neighbourhood of $w_0$ in the simplex interior and that the within-layer Fisher information in a local simplex chart is positive definite. Then, conditional on $\eta_0$, for any sequence $M_P\to\infty$,
\[
\Pi\left(\|W_\ell-w_0\|_2>\frac{M_P}{\sqrt{P}} \,\middle|\, D_\ell,\eta_0\right)\to 0
\quad\text{as } P\to\infty.
\]
\end{theorem}

\section{Numerical study}
\label{sec:numerical}

Building upon the theoretical properties in Section~\ref{sec:theory}, we conduct a simulation study to evaluate BALM against existing methodologies. The experiments are designed to isolate the two core methodological departures of BALM: the ability of continuous mixed-membership representations to capture overlapping network topologies compared to hard-clustering block models, and the inferential benefits of explicitly modelling structure-informed sparsity through the hurdle coupling parameter.

\subsection{Simulation design and evaluation metrics}
\label{sec:simulation_design}

We generate continuous, zero-inflated multilayer networks with $M = 3$ diffuse latent score templates $S_m = \tau U_m \diag(\gamma_m) U_m^\top$ and orthonormal matrices $U_m \in \mathbb{R}^{n \times K}$. Subject-specific heterogeneity is introduced via Dirichlet mixing weights $W_\ell \sim \mathrm{Dir}\{(\alpha/M) \mathbf{1}_M\}$, where $\alpha \in \{0.1, 0.3\}$ simulates near-hard clustering and $\alpha \in \{1.0, 3.0\}$ produces overlapping memberships. The latent mean is $\mu_\ell = \sum_{m=1}^M W_{\ell m} \offdiag(S_m)$. Edge existence follows a Bernoulli distribution with probability $\sigmoid(a_0^\star + a_1^\star \mu_{\ell, ij})$, and conditional intensities are drawn from a Gaussian centred at $\mu_{\ell, ij}$ with variance $\sigma^2$. We set $a_0^\star = -2.5$ and $a_1^\star = 5.0$, and randomly mask 15\% of edges for out-of-sample evaluation.

Performance is assessed using five metrics on the held-out test set: template correlation, link-prediction area under the receiver-operating-characteristic curve (AUC), continuous-weight mean squared error (MSE), relative Frobenius error, and adjusted Rand index (ARI) computed from the dominant mixing weight. Topology metrics serve as the fairest common comparison across all methods, whereas weight-channel metrics are most meaningful for BALM and other models that explicitly target $Y\mid Z=1$.

\subsection{Mixed-membership versus hard-clustering architectures}
\label{sec:exp_mixed_membership}

\begin{figure}[htbp]
    \centering
    \includegraphics[width=0.95\textwidth]{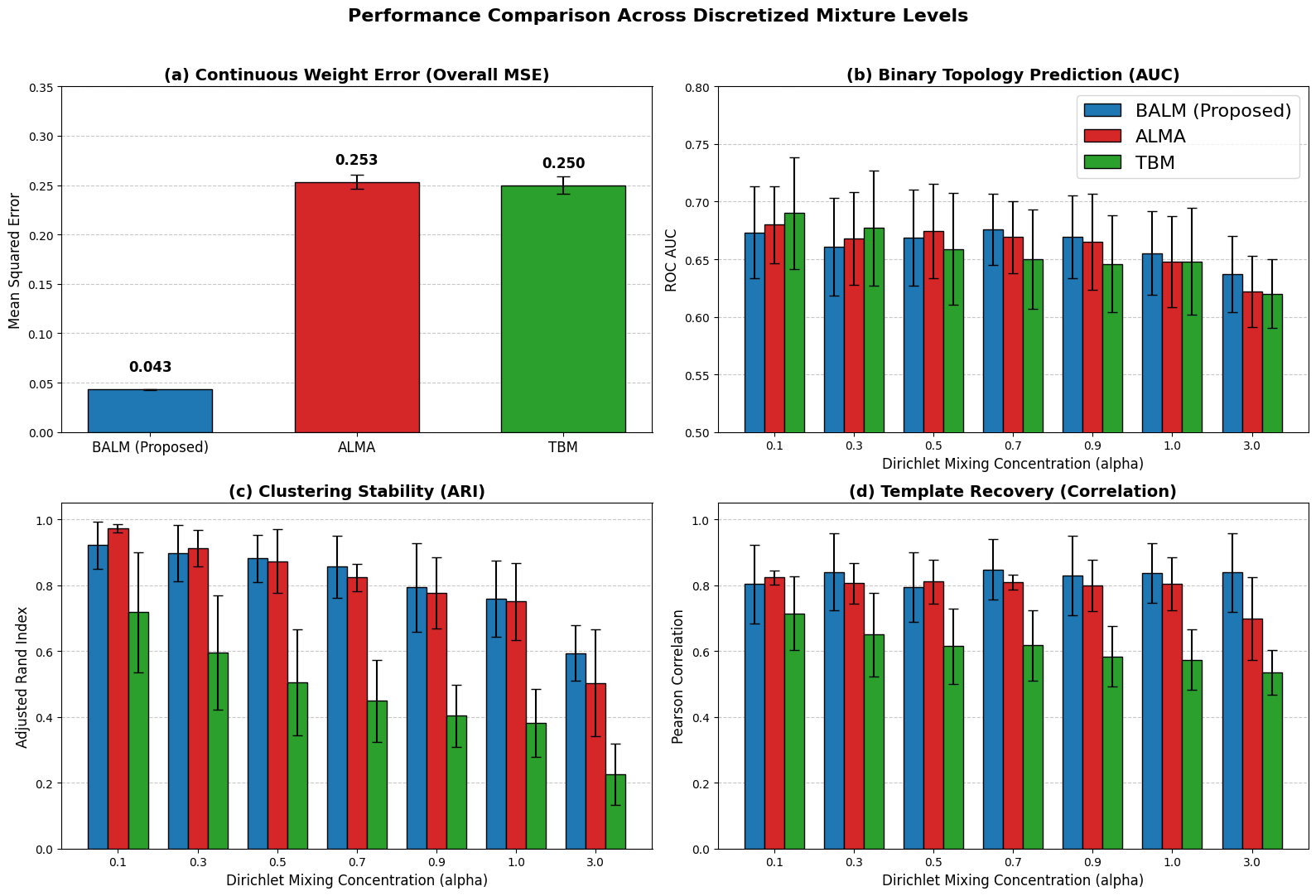}
    \caption{Performance comparison between BALM, ALMA, and the tensor block model across varying Dirichlet concentration parameters ($\alpha$). All reported metrics are averaged over 50 independent replications per setting. Panels show continuous-weight mean squared error, binary topology prediction, clustering stability and template recovery.}
    \label{fig:sim_mixed}
    \par\smallskip\noindent\textbf{Alt text:} Four-panel bar chart comparing BALM, ALMA and the tensor block model across increasing Dirichlet concentration. The panels report continuous-weight error, binary topology prediction, adjusted Rand index and template correlation.
\end{figure}

This experiment evaluates parameter recovery and out-of-sample prediction across varying degrees of layer-level mixed membership. The Dirichlet concentration parameter $\alpha$ controls the extent of this mixing: smaller values generate near-discrete layer assignments resembling hard clustering, while larger values produce overlapping convex combinations of multiple latent templates. Figure~\ref{fig:sim_mixed} compares BALM against ALMA and the tensor block model across these conditions. Detailed numerical results across all metrics are provided in Appendix D of the Supplementary Material.

In the near-discrete scenario, the data generation process inherently favours discrete class assignments. Under these specific conditions, the deterministic baseline ALMA yields the highest adjusted Rand index and slightly stronger template correlation. BALM achieves comparable, though sometimes slightly lower, scores in this sparse setting, as its continuous Dirichlet mixture parameterisation introduces modest estimation variance when the true underlying structure is close to categorical. The tensor block model yields lower adjusted Rand index and correlation across all settings, reflecting the limitations of blockwise-constant mean constraints when the data are generated from diffuse low-rank templates.

As the mixing level increases, the assumption of mutually exclusive categories diverges from the data-generating mechanism. Consequently, the performance of hard-clustering baselines declines. At $\alpha=3.0$, ALMA's adjusted Rand index decreases to $0.5028$ and the tensor block model's to $0.2260$, whereas BALM records $0.5938$ and preserves a template correlation of $0.8384$. These differences are anticipated from the simplex parameterisation discussed in Section~\ref{sec:model}, which robustly handles overlapping memberships.

For topology prediction, all models perform comparably within a narrow band. For weight recovery, BALM substantially outperforms the topology-only baselines, as expected, because it explicitly models the nonzero-weight channel. We therefore regard topology prediction, adjusted Rand index and template correlation as the primary cross-method evidence, while the mean squared error results demonstrate the advantage of modelling the continuous channel directly.

\subsection{The impact of structure-informed sparsity}
\label{sec:exp_sparsity_coupling}

To evaluate the structure-informed coupling mechanism introduced in Section~\ref{sec:model}, we simulate populations of $L=100$ replicated networks over $n=68$ nodes using $M=3$ templates of rank $K=4$. We inject substantial Gaussian noise ($\sigma=1.0$) into the continuous logit-scale edge weights, forcing models to leverage topological information. To isolate the coupling effect from overall density changes, we numerically calibrate the baseline intercept $a_0^\star$ so that the marginal expected network density remains fixed at 15\%, 30\%, or 50\% across true coupling strengths $a_1^\star \in \{0.0, 1.0, 2.0, 3.0, 4.0, 5.0\}$.

\begin{figure}[htbp]
    \centering
    \includegraphics[width=0.95\textwidth]{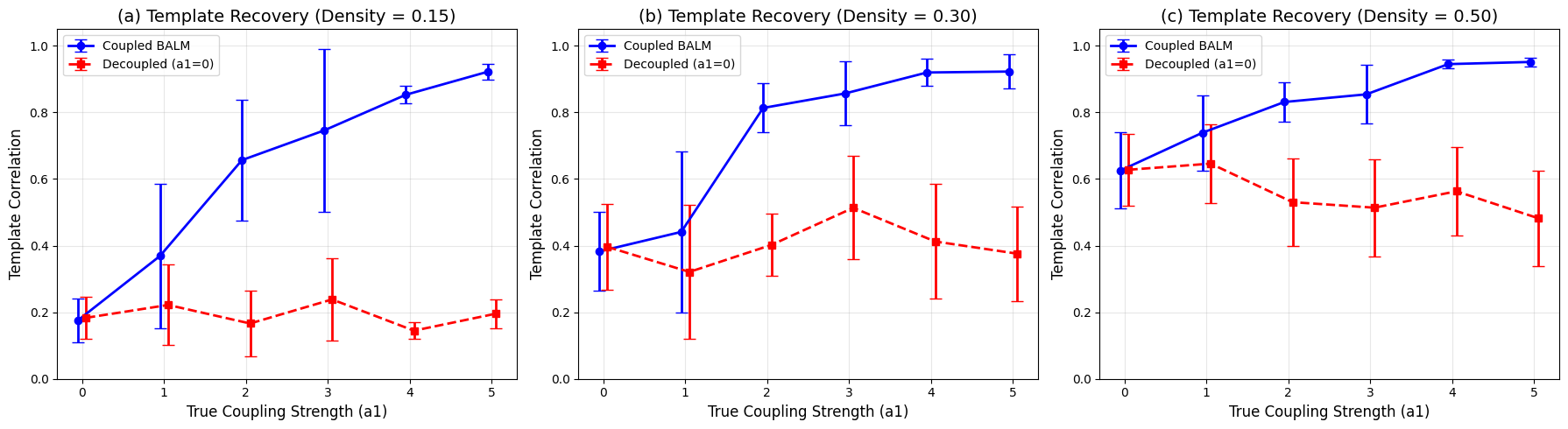}
    \caption{Performance comparison illustrating the effect of structure-informed sparsity. Panels show template correlation for the coupled BALM and restricted decoupled models across true coupling strengths at 15\%, 30\%, and 50\% target network densities. Error bars represent standard deviations over 5 replications.}
    \label{fig:sparsity_coupling_plot}
    \par\smallskip\noindent\textbf{Alt text:} Three-panel line plot showing template-recovery correlation as true coupling strength increases. The coupled model increases with coupling strength, while the decoupled model remains lower, especially in sparse networks.
\end{figure}

Under these calibrated conditions, we compare a coupled BALM model that jointly infers the latent templates and the coupling parameter $a_1$ with a restricted decoupled model forced to assume $a_1=0$. Figure~\ref{fig:sparsity_coupling_plot} summarises the template-recovery performance, with the full quantitative breakdown provided in Appendix D of the Supplementary Material.

\section{Human Connectome Project brain connectome data}
\label{sec:hcp}

To demonstrate the empirical utility of BALM, we analyse brain connectomes derived from the Human Connectome Project S1200 release. The statistical goal is to evaluate whether BALM's mixed-membership formulation and hurdle likelihood recover stable latent network topologies in a large structural connectome cohort, and whether those latent topologies admit plausible exploratory annotation using behavioural variables. We compare this behaviour with dimension-matched deterministic clustering approaches, while avoiding claims that would require confirmatory external validation.

\subsection{Structural connectomics: specification and robust shrinkage}
\label{sec:hcp_structural_spec}

We analyse structural connectomes for a subset of $L=1065$ subjects using the 68-node Desikan--Killiany atlas, yielding $P=2278$ undirected edges per subject. The structural matrices are sparse and highly heterogeneous in their nonzero weights. The weights were normalised to lie in $(0,1)$ for nonzero entries, and exact zeros were retained as structural absences for the hurdle formulation. Because the empirical distribution of nonzero logit-scale weights is markedly heavy tailed, we fit BALM using a Student-$t$ likelihood with $\nu=5$ for the continuous channel.

The subject inclusion criteria, tractography workflow, parcellation registration procedure, edge-weight definition, normalisation rule, behavioural-variable missingness handling and train-test split seeds are documented in the reproducibility repository described in the Code availability statement. The current analysis treats exact zero entries as structural absences after preprocessing and models strictly positive entries through the conditional continuous channel.

After aligning the templates via the Hungarian algorithm to resolve label switching, the between-run differences were negligible and concentrated primarily in low-strength peripheral connections rather than in the large-scale topology of the dominant motifs. This structural reproducibility suggested that the recovered connectivity patterns reflect stable population characteristics rather than idiosyncratic local posterior modes.

Posterior predictive checks (detailed in Appendix E of the Supplementary Material) support the use of robust shrinkage. The empirical density of nonzero logit-scale edge weights is strongly leptokurtic. A Gaussian specification spreads mass too broadly in order to accommodate large structural hubs, whereas the Student-$t$ likelihood is able to follow both the sharp central peak and the tail behaviour more closely. Across candidate values of $M$, the Student-$t$ model with $M=5$ templates achieved the best combination of the information criterion, held-out topology prediction and between-run template stability, and we treat that fit as the primary structural representation.

\subsection{Disentangling overlapping circuitry in structural connectomes}
\label{sec:hcp_disentanglement}

\begin{figure}[htbp]
    \centering
    \includegraphics[width=0.95\textwidth]{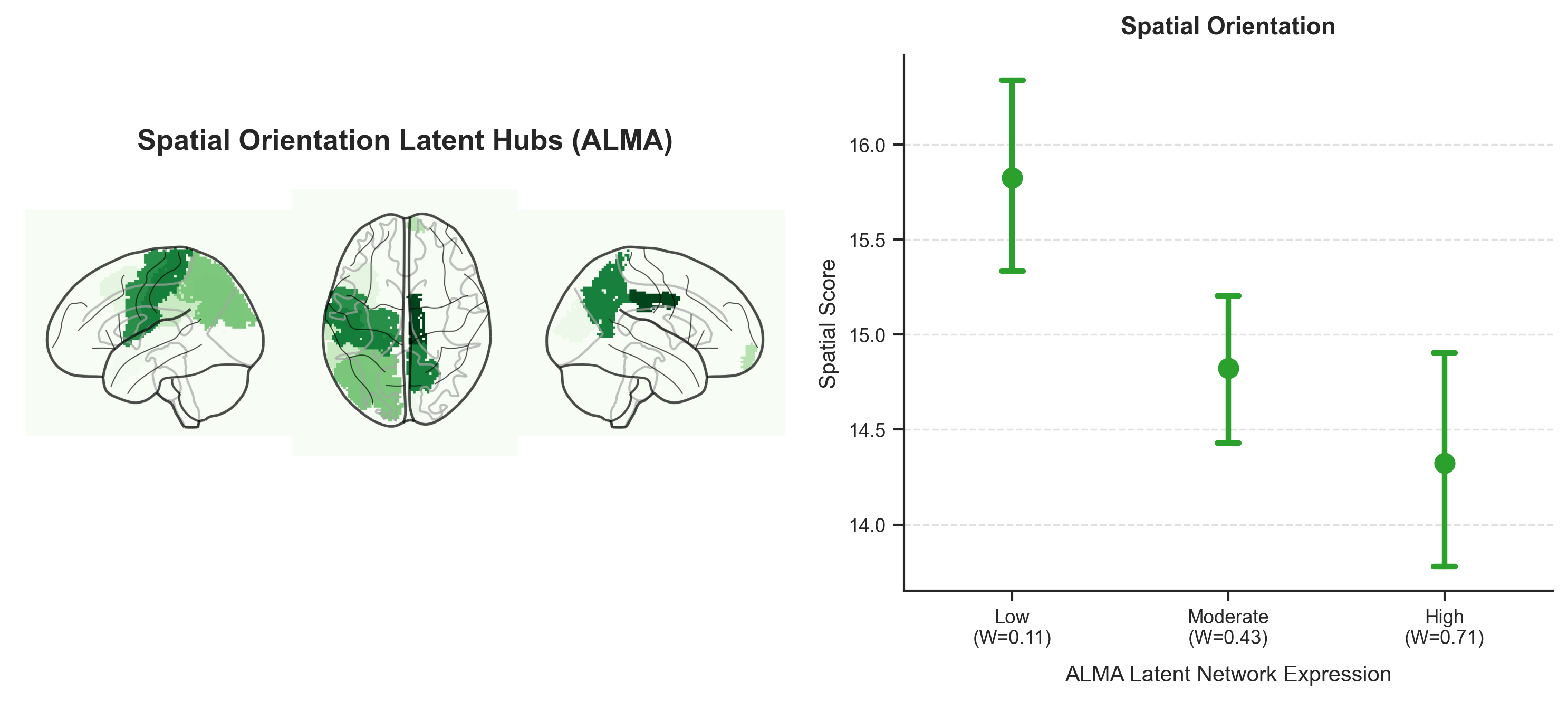}
    \caption{Representative latent hub derived from ALMA. Under orthogonality constraints, the component merges default-mode and somatomotor regions into a single mixed template.}
    \label{fig:alma_behavior}
    \par\smallskip\noindent\textbf{Alt text:} Brain-plot and behavioural-summary figure for an ALMA component, showing default-mode and somatomotor involvement and a descriptive association with Spatial Orientation.
\end{figure}

A central question in connectomics is whether the latent representation separates overlapping large-scale systems or instead compresses them into broad composite factors. To assess this, we compared BALM with a dimension-matched ALMA implementation using $M=5$ latent components. Because ALMA imposes orthogonality constraints on its subject-level representation, it tends to consolidate overlapping systems into singular latent variables. In our analysis, one representative ALMA template was dominated jointly by higher-order default-mode and somatomotor regions, producing an anatomically mixed component that was difficult to interpret as a single coherent system (Figure~\ref{fig:alma_behavior}).

BALM, by contrast, allows subject-level expression to vary through simplex weights without orthogonality constraints. The resulting structural templates are still broad, as expected in highly sparse dMRI data, but they are more easily interpreted as dominant motifs with graded overlap rather than as forced amalgamations. The template structure is stable across repeated runs after alignment and gives a more nuanced picture of subject heterogeneity.

\begin{figure}[htbp]
    \centering
    \includegraphics[width=0.95\textwidth]{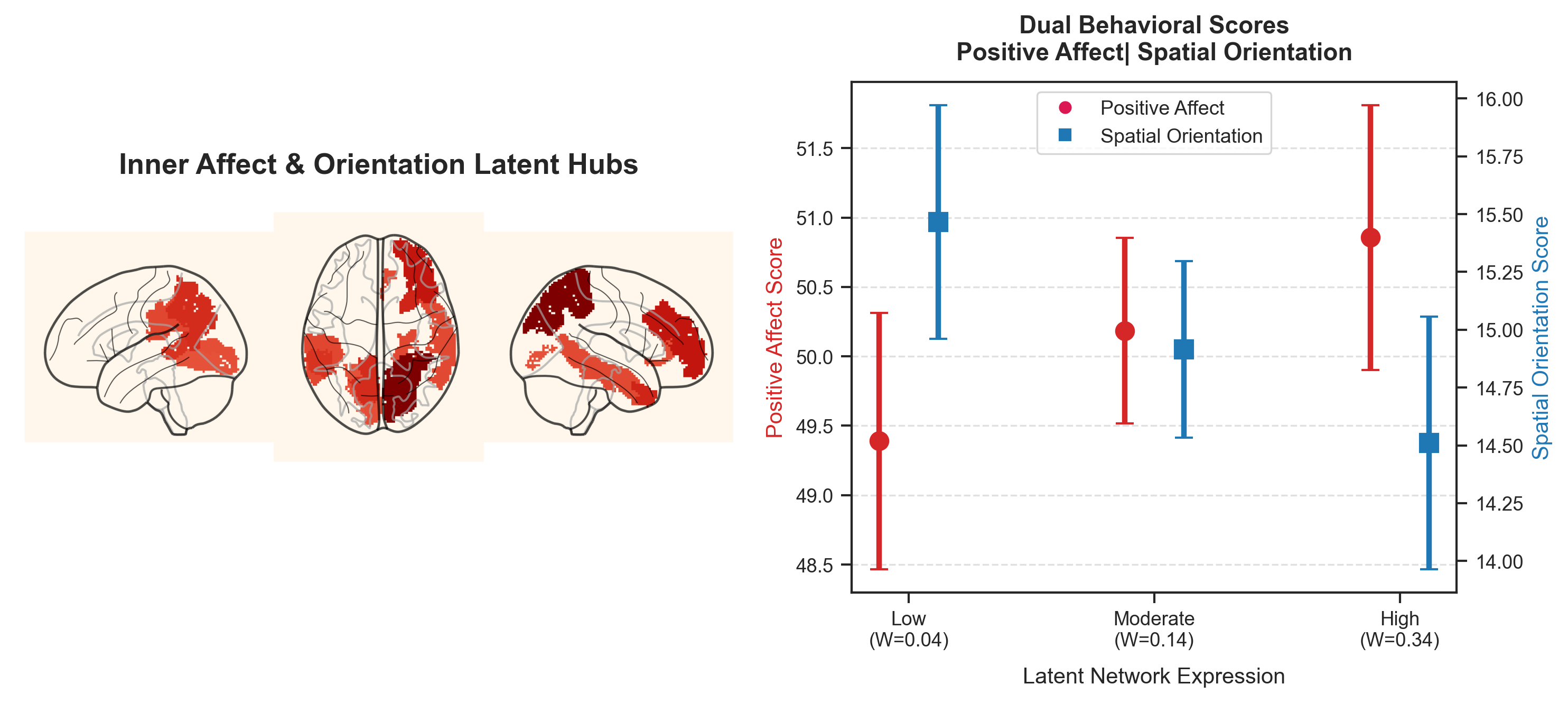}
    \caption{Representative BALM structural template dominated by default-mode regions. The accompanying behavioural associations are exploratory annotations and are reported descriptively rather than as confirmatory findings.}
    \label{fig:dmn_behavior}
    \par\smallskip\noindent\textbf{Alt text:} Brain-plot and behavioural-summary figure for a BALM structural template dominated by default-mode regions, with descriptive Positive Affect and Spatial Orientation summaries across expression groups.
\end{figure}

\begin{figure}[htbp]
    \centering
    \includegraphics[width=0.95\textwidth]{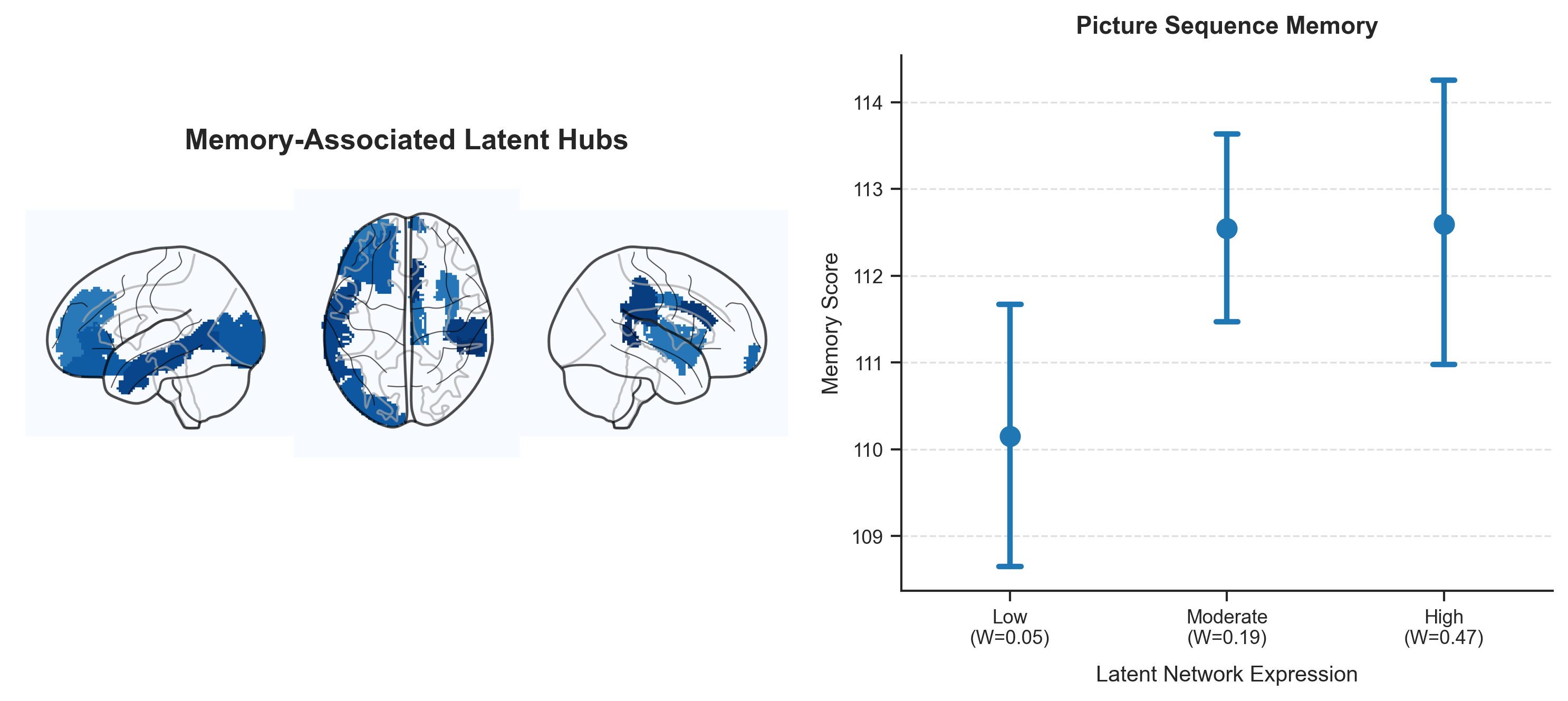}
    \caption{Representative BALM structural template with default-mode and limbic involvement. The small negative association with Picture Sequence Memory is used as exploratory annotation only.}
    \label{fig:fpn_memory}
    \par\smallskip\noindent\textbf{Alt text:} Brain-plot and behavioural-summary figure for a BALM structural template with default-mode and limbic involvement, with Picture Sequence Memory shown descriptively across expression groups.
\end{figure}

The divergence between ALMA's composite templates and BALM's isolated motifs highlights the practical impact of the model choices described in Section~\ref{sec:model}. As illustrated by a direct visual comparison between Figure~\ref{fig:alma_behavior} and Figures~\ref{fig:dmn_behavior}--\ref{fig:fpn_memory}, the orthogonality constraint forces ALMA to agglomerate distinct networks into a single dense hub. In contrast, BALM disentangles overlapping but distinct circuits---such as core default-mode circuitry and limbic-associated pathways---into separate templates. By avoiding artificial topological convolution, BALM provides a more plausible approximation of the graded functional organisation of the human cortex.

We used behavioural measures only as exploratory annotations of the recovered structural templates. The observed correlations are small in magnitude and should not be interpreted as confirmatory evidence for a mechanistic brain--behaviour relation. Instead, they serve as a descriptive check on whether templates align with recognisable cognitive or affective axes. In this descriptive sense, one default-mode dominated structural template (Figure~\ref{fig:dmn_behavior}) was positively associated with Positive Affect and negatively associated with Spatial Orientation, while another template involving default-mode and limbic regions (Figure~\ref{fig:fpn_memory}) showed a mild negative association with Picture Sequence Memory. The primary statistical result is the stability and heterogeneity of the latent structural representation rather than the magnitude of the behavioural effect sizes.

While our primary focus is structural connectomics, Appendix E of the Supplementary Material shows that BALM can also be applied to dense functional connectomes, where recovered templates can be annotated descriptively with behavioural summaries.

\section{Discussion}
\label{sec:discussion}

This work introduced BALM, a generative Bayesian framework for sparse multilayer networks that explicitly decouples edge existence from conditional edge strength via a hurdle likelihood. By representing shared structure through simplex-mixed low-rank score templates, BALM moves beyond discrete clustering to quantify graded, overlapping subject-specific expression. In structural connectomics, where preprocessing often induces structural zeros, this joint modelling prevents conflating topological absence with weak connectivity. Our application to the Human Connectome Project illustrates these advantages relative to restrictive orthogonal clustering approaches like ALMA. Rather than artificially amalgamating distinct brain systems, BALM recovers stable, continuous template weights that yield nuanced summaries of population heterogeneity, accompanied by exploratory behavioural annotations.

Several assumptions and practical considerations guide the current framework. The logit transformation requires normalised nonzero weights in $(0,1)$, and we assume conditional independence of edges given the latent templates to maintain scalable likelihood evaluation without explicitly modelling residual transitivity. In practice, the template rank $K$ should be chosen for interpretability, while the number of templates $M$ is guided by information criteria and consensus reproducibility. Furthermore, our asymptotic guarantees focus on an inferential scenario with fixed template complexity and increasing numbers of replicated networks.

These model properties point to several natural extensions. The simplex mixing weights $W_\ell$ can be expanded into flexible regressions to allow direct inference on how demographic or clinical covariates modulate template expression. Additionally, one could accommodate alternative heavy-tailed weight distributions, or introduce nonparametric mixture priors to infer $M$ automatically and discourage redundant templates. Finally, scalable computational approximations, such as variational inference or subsampling-based Markov chain Monte Carlo, would further extend BALM to high-resolution node sets or longitudinal studies with massive repeated measurements.

\section*{Data availability}

The imaging and behavioural data used in this work were obtained from the Human Connectome Project (HCP) Young Adult S1200 release through ConnectomeDB. Access to these raw data requires registration and acceptance of the applicable HCP data-use terms at \url{https://db.humanconnectome.org}. Open-access HCP data include the imaging pipelines and most behavioural measures, whereas restricted variables require separate institutional approval. Subject-level restricted variables and raw MRI data are not redistributed with this article. Where permitted by HCP data-use terms, the reproducibility repository provides the subject-inclusion manifest, non-restricted derived summaries and simulated data-generating scripts necessary to reproduce the analyses.

\section*{Code availability}

Code to reproduce all numerical experiments and empirical analyses, including the Hamiltonian Monte Carlo sampling framework for the Bayesian Adaptive Latent Mixture (BALM) model, is publicly available on GitHub at \url{https://github.com/JackieChenYH/BALM}. The repository includes the \texttt{experiments} folder containing the simulation pipelines, and the \texttt{HCP} folder providing standalone scripts for the model formulation, fitting workflow and posterior diagnostics for both sparse structural and dense functional networks. A citable archived release of this repository with a persistent DOI will be generated and added to this statement prior to final acceptance.

\section*{Conflict of interest}

The authors declare no competing interests.

\section*{Author contributions}

Hsin-Hsiung Huang: conceptualisation, methodology, theoretical development, investigation, writing--original draft, writing--review and editing, supervision, and funding acquisition. Yuh-Haur Chen: algorithmic implementation and software development, design and execution of numerical experiments, neuroimaging data curation, formal validation, visualisation, and writing--review and editing. Teng Zhang: conceptualisation, methodology, theoretical development, investigation, writing--review and editing, supervision, and funding acquisition. All authors reviewed and approved the final manuscript.

\section*{Funding}

This work was supported in part by the U.S. National Science Foundation under grants DMS-1924792 and DMS-2318925 to Hsin-Hsiung Huang and CNS-1818500 to Teng Zhang. The funders had no role in the design of the study, the analysis or interpretation of the results, or the writing of the manuscript.

\section*{Acknowledgements}

Data used in this work were provided by the Human Connectome Project, WU-Minn Consortium. The authors thank the Human Connectome Project investigators and participants for making these data available to the research community. The authors also thank colleagues and seminar participants who provided comments on earlier versions of this work. A generative artificial intelligence tool was used for editorial and LaTeX-formatting checks during manuscript preparation. All scientific content, analyses, code and final wording were reviewed and verified by the authors.

\setlength{\bibsep}{0pt}
\bibliographystyle{abbrvnat}
\bibliography{references}

\end{document}